\documentclass{ptephy_v1}

\preprintnumber{XXXX-XXXX} 
\usepackage{hyperref}

\usepackage{graphicx}
\usepackage{dcolumn}
\usepackage{bm}
\usepackage{mathrsfs}
\usepackage{natbib}
\usepackage{subcaption}
\usepackage{multirow}
\bibpunct{(}{)}{;}{a}{,}{,}

\begin{document}

\title{
    A Self-regulated Stochastic Acceleration Model of Pulsar Wind Nebulae
}


\author[1,2]{Shuta J. Tanaka}

\author[3,4,5]{Wataru Ishizaki}

\affil[1]{Department of Physical Sciences, Aoyama Gakuin University, 5-10-1 Fuchinobe, Sagamihara Kanagawa 252-5258, Japan
\email{sjtanaka@phys.aoyama.ac.jp}
}

\affil[2]{Graduate School of Engineering, Osaka University, 2-1 Yamadaoka, Suita, Osaka 565-0871, Japan}

\affil[3]{Yukawa Institute for Theoretical Physics, Kyoto University, Kitashirakawa-Oiwake-cho, Sakyo-ku, Kyoto 606-8502, Japan}
\affil[4]{Astronomical Institute, Graduate School of Science, Tohoku University, Sendai 980-8578, Japan}
\affil[5]{Frontier Research Institute for Interdisciplinary Sciences, Tohoku University, Sendai 980-8578, Japan}



\begin{abstract}%
    Pulsar wind nebulae (PWNe) are clouds of the magnetized relativistic electron/positron plasma supplied from the central pulsar.
    However, the number of radio-emitting particles inside a PWN is larger than the expectation from the study of pulsar magnetospheres and then their origin is still unclear.
    A stochastic acceleration of externally injected particles by a turbulence inside the PWN is proposed by our previous studies.
    In this paper, the previous stochastic acceleration model of the PWN broadband spectra is improved by taking into account the time evolution of the turbulent energy and then the total energy balance inside a PWN is maintained.
    The turbulent energy supplied from the central pulsar is wasted by the backreaction from the stochastic particle acceleration and the adiabatic cooling according the PWN expansion.
    The model is applied to the Crab Nebula and reproduce the current broadband emission spectrum, especially the flat radio spectrum although time evolution of the turbulent energy (diffusion coefficient) is a bit complicated compared with our previous studies, where we assumed an exponential behavior of the diffusion coefficient.
\end{abstract}

\subjectindex{xxxx, xxx}

\maketitle

\section{Introduction}\label{sec:intro}

Pulsars are powerful wind-blowing objects that form relativistic magnetized plasma bubbles called pulsar wind nebulae (PWNe).
PWNe are expanding with time and a typical size of a few pc at an age of a few kyr (c.f. Ref. \cite{Gaensler&Slane06}).
Broadband emission from radio through PeV gamma-rays is produced by high-energy particles accelerated at the termination shock of the relativistic pulsar wind \cite{LHAASOCrab21,LHAASOPeVatron21}.
Dynamics of relativistic magnetized plasmas, particle acceleration mechanism, and the properties of their central pulsars can be studied from the PWNe \cite{Porth+17, Reynolds+17}.

One of the well-known problems for the dynamics of the PWN is the $\sigma$-problem, where the magnetization of the pulsar wind at the termination shock is required to be much smaller than unity (c.f. Refs. \cite{Rees&Gunn74, Kennel&Coroniti84a}).
The three-dimensional magnetohydrodynamic simulation \cite{Porth+14} pointed out that a possible answer to the $\sigma$-problem is `turbulence' inside the PWN (c.f. Refs. \cite{Zrake&Arons17, Tanaka+18}).
The turbulence would also play a role in particle `spatial' diffusion \cite{Porth+16, Ishizaki+17, Ishizaki+18, vanRensburg+20}, where the observed extent of PWNe in X-rays \cite{Bamba+10} and in $\gamma$-rays \cite{HAWC17} can be explained by spatial diffusion of the accelerated particles.
Polarization observations \cite{Bietenholz&Kronberg91, Aumont+10, Bucciantini+23} also require the turbulent component of the magnetic field comparable with the ordered magnetic field \cite{Bucciantini+05, Nakamura&Shibata07, Bucciantini+17}.

In this paper, we study the stochastic particle acceleration by the turbulence inside the PWN (particle diffusion in momentum space).
Tanaka and Asano (2017, hereafter TA17) \cite{Tanaka&Asano17} developed a stochastic acceleration model of PWNe in order to explain the origin of the low-energy ($\gamma \lesssim 10^6$) radio-emitting particles which are thought as a distinct population from the high-energy ($\gamma \gtrsim 10^6$) X-ray-emitting particles injected from the central pulsar (c.f. Refs. \cite{Kennel&Coroniti84b, Atoyan&Aharonian96}).
The observed amount of the radio-emitting particles are much more than the central pulsar can produce in its life time \cite{Tanaka&Takahara10, Tanaka&Takahara11, Tanaka&Takahara13a, Timokhin&Harding15}.
The radio-emitting particles would be served externally, such as the surrounding supernova ejecta for example, and they are accelerated stochastically to form the observed hard radio spectra \cite{Tanaka&Asano17}.
However, in their model, the total energy was not conserved because the origin of the turbulence was not taken into account \cite{Tanaka&Asano17, Tanaka&Kashiyama23}.

Here, we develop a self-regulated turbulent acceleration model of PWNe.
We solve the evolution of the turbulent energy $E_{\rm T}(t)$ regarding it as a component of the PWN.
The turbulent energy is continuously supplied from the central pulsar and is subsequently converted into the energy of the radio-emitting particles by the stochastic acceleration process (e.g. Ref. \cite{Kakuwa16}).
In Section \ref{sec:Model}, we describe our self-regulated acceleration model, and a relation with the diffusion coefficient in the momentum space and the energy of the turbulence is introduced.
In Section \ref{sec:Results}, the results for the present model applying to the Crab Nebula are shown.
We summarize the present paper in Section \ref{sec:Dis&Cons}.

\section{Self-regulated Stochastic Acceleration Model}\label{sec:Model}

%
\begin{table}
\caption{
    Summary of the model parameters for Figs. \ref{fig:IMP} and \ref{fig:CONT}.
}
    \label{tbl:Parameters}
    \begin{minipage}{0.5\textwidth}
        \centering
        \begin{tabular}{ccc}
        \hline
        \multicolumn{3}{c}{Fitted Parameters} \\
        \hline
	    Symbol             & Impulsive            & Continuous \\
        \hline
        $\tau_{\rm acc}$   & \multicolumn{2}{c}{10~{\rm yr}}             \\ 
        $f_{\rm imp}$      & $2.5 \times 10^{-2}$ & $\cdots$             \\ 
	    $f_{\rm cont}$     & $\cdots$             & $8.0 \times 10^{-6}$ \\ 
    	$\eta_{\rm B}$     & 3.0 $\times 10^{-3}$ &	5.0 $\times 10^{-3}$ \\ 
        $\eta_{\rm T}$     & 0.8                  & 0.4                  \\ 
        $\eta_{\rm e}$     & \multicolumn{2}{c}{0.2}                     
        \end{tabular}
    \end{minipage}
    \begin{minipage}{0.5\textwidth}
        \centering
        \begin{tabular}{cccc} 
        \hline
        \multicolumn{4}{c}{Fixed Parameters}  \\
        \hline
	    $E_{\rm SN}$       & $10^{51}$ erg   &	$M_{\rm ej}$       & 9.5 $M_{\odot}$  \\
	    $n_{\rm ISM}$      & 0.1 cm$^{-3}$   &  $d$                & 2 kpc   \\
	    $M_{\rm NS}$       & 1.4 $M_{\odot}$ &	$R_{\rm NS}$       & 12 km   \\
	    $P_0$              & 18.7 msec       &	$B_0$              & $3.35 \times 10^{12}$ G \\
	    $n$                & 2.51            &	$p$                & 2.4     \\
        $\gamma_{\rm min}$ & $10^{6}$        & $\gamma_{\rm max}$  & 7.0 $\times 10^{9}$  \\
        $\gamma_{\rm inj}$ & $1.26$          & $t_{\rm init}$      & 1 yr
        \end{tabular}
    \end{minipage}
\end{table}

Here, the self-regulated stochastic acceleration model is described focusing on improvements from the previous studies \cite{Tanaka&Asano17, Tanaka&Kashiyama23}.
The turbulence inside the PWN ($E_{\rm T}(t)$) is explicitly set to an element in addition to the magnetic field ($E_{\rm B}(t)$) and the relativistic electrons/positrons ($E_{\rm e}(t)$).
We introduce the fraction parameters, $\eta_{\rm T}$, $\eta_{\rm B}$ and $\eta_{\rm e}$, where they control the injection rate of the turbulent $\dot{E}_{\rm T}$, magnetic $\dot{E}_{\rm B}$ and particle (electron/positron) $\dot{E}_{\rm e}$ energy from the central pulsar.
The spin-down power of the central pulsar ($L_{\rm spin}(t)$) is divided into the three components, such as $\dot{E}_i = \eta_i L_{\rm spin}~(i = {\rm T, B, e})$, and they satisfy the condition,
\begin{eqnarray}\label{eq:FractionParameters}
    \eta_{\rm T} + \eta_{\rm B} + \eta_{\rm e} \le 1,
\end{eqnarray}
where we expect $\eta_{\rm T}$ is the same order as $\eta_{\rm e}$, while $\eta_{\rm B} \ll 1$.
In the present model, we do not require the sum of a fraction parameters (Eq. (\ref{eq:FractionParameters})) to be unity, because $\eta_{\rm T}$ is attributed to acceleration only for electrons while the ions injected externally (see Eq. (\ref{eq:ExternalInjection})) would also be accelerated (see discussion in Section \ref{sec:Dis&Cons}).
The turbulent energy interacts with ions are ignored because the emission from the accelerated ions are inefficient.

First, one of the roles of the turbulence within the one-zone spectral evolution model is a particle diffusion in the momentum space.
Introducing the time-scale of the stochastic acceleration $t_{\rm acc}$, the diffusion coefficient is written as
\begin{eqnarray}\label{eq:DiffusionCoefficient}
    D_{\gamma \gamma}(\gamma, t) = \frac{\gamma^2}{2 t_{\rm acc}(\gamma, t)}
\end{eqnarray}
where $\gamma$ is the Lorentz factor of electrons/positrons and $t_{\rm acc}$ depends both on $\gamma$ and $t$ in general.
Based on the quasilinear approximation of the gyroresonant scattering off the particles by turbulence, this acceleration time is inversely proportional to the spectral energy of the turbulence whose wavenumber corresponds the gyro-radius of an electron of $\gamma$ (e.g. Refs. \cite{Schlickeiser02, Kakuwa16}).
Nevertheless, here, we use the hardsphere formula assuming the non-resonant scattering between the turbulence, i.e., $t_{\rm acc}$ is independent from $\gamma$ according to Tanaka and Kashiyama (2023, hereafter TK23 \cite{Tanaka&Kashiyama23}).
The acceleration time is still expected to be inversely proportional to the energy of the spectrally integrated turbulent energy $E_{\rm T}(t)$, and then we introduce
\begin{eqnarray}\label{eq:AccelerationTime}
    t_{\rm acc}(t) = \tau_{\rm acc} \frac{\eta_{\rm T} E_{\rm rot}(t)}{E_{\rm T}(t)},
\end{eqnarray}
where $E_{\rm rot}(t) = \int L_{\rm spin} dt$ is the total injected rotational energy from the central pulsar.
In the present model, the acceleration time-scale $t_{\rm acc}(t)$ changes with time because of the decay mechanisms of the turbulence discussed below, i.e., we consider the case $E_{\rm T}(t) \ne \eta_{\rm T} E_{\rm rot}(t)$.
$\tau_{\rm acc}$ is the acceleration time-scale without any decay processes of the turbulence and is an only parameter that controls the stochastic acceleration.
Eq. (\ref{eq:AccelerationTime}) is a major difference from the previous studies, where an artificial parameter of the turbulence decay time-scale was implemented (see Eq. (7) of TK23).
We wiil discuss validity of Eq. (\ref{eq:AccelerationTime}) in section \ref{sec:Dis&Cons}.

The evolution of the turbulent energy $E_{\rm T}(t)$ would be expressed as
\begin{eqnarray}\label{eq:TurbulentEnergyEvolution}
	\frac{d E_{\rm T}}{dt} 
	& = &
    \eta_{\rm T} L_{\rm spin} 
    -
    \left(\frac{\delta E_{\rm T}}{\delta t}\right)_{\rm damp}
    -
    \frac{E_{\rm T}}{t_{\rm adi}(t)},
\end{eqnarray}
where $t_{\rm adi}(t) \equiv R_{\rm PWN}(t) / \dot{R}_{\rm PWN}(t)$ is the adiabatic cooling time-scale and $R_{\rm PWN}$ is the radius of the PWN.
The first term in the right-hand side of Eq. (\ref{eq:TurbulentEnergyEvolution}) is injection from the pulsar as a random bulk motion of electron/positron plasma as a magnetohydodynamic fluid.
The second term represents that the turbulence is attenuated by the backreaction of the stochastic acceleration of electrons/positrons \cite{Kakuwa16} and will be defined in Eq. (\ref{eq:BackReaction}).
Finally, the third term is the adiabatic cooling by the expansion of the PWN \cite{Robertson&Goldreich12}.
Note that the turbulence should have the pressure, which we assume one-third of the turbulent energy density and the turbulent pressure is taken into account for the expansion of the PWN $R_{\rm PWN}(t)$ in addition to the particle and magnetic pressure (c.f. Refs. \cite{Gelfand+09, Bandiera+20, Tanaka&Kashiyama23}).

Eq. (\ref{eq:TurbulentEnergyEvolution}) is solved with the energy distribution of the accelerated particles $N(\gamma, t)$,
\begin{eqnarray}\label{eq:FokkerPlanckEquation}
    \frac{\partial}{\partial t} N
    +
    \frac{\partial}{\partial \gamma} \left[\left(\dot{\gamma}_{\mathrm{cool}} - \gamma^{2} D_{\gamma \gamma} \frac{\partial}{\partial \gamma} \frac{1}{\gamma^{2}}\right) N \right] 
    =
    Q_{\rm ext}(\gamma, t)
    +
    Q_{\rm PSR}(\gamma, t),
\end{eqnarray} 
where $\dot{\gamma}_{\rm cool}$ includes the adiabatic, synchrotron and inverse Compton coolings, $Q_{\rm ext}$ represents an external source of the radio-emitting particles and $Q_{\rm PSR}$ is the (single power-law) particle injection from the pulsar.
The total particle energy inside the PWN is given by $E_{\rm e}(t) = \int d \gamma \gamma m_{\rm e} c^2 N(\gamma, t)$, where $m_{\rm e}$ and $c$ are the electron mass and the speed of light, respectively.
The third term on the left-hand side of Eq. (\ref{eq:FokkerPlanckEquation}) is the interaction between the turbulence and particles, and relates with the damping term in Eq. (\ref{eq:TurbulentEnergyEvolution}) as
\begin{eqnarray}\label{eq:BackReaction}
    \left(\frac{\delta E_{\rm T}}{\delta t}\right)_{\rm damp}
    \equiv
    \int d \gamma \gamma m_{\rm e} c^2
    \frac{\partial}{\partial \gamma} 
    \left( - \gamma^2 D_{\gamma \gamma} \frac{\partial}{\partial \gamma} \frac{N}{\gamma^2} \right),
\end{eqnarray}
The backreaction of the stochastic particle acceleration to the turbulence is formulated self-consistently and the acceleration is suppressed by the decay of turbulence.

The external particle injection is described as
\begin{eqnarray}\label{eq:ExternalInjection}
	Q_{\rm ext}(\gamma, t) =
	\left\{
   	\begin{array}{ll}
        f_{\rm imp} (E_{\rm rot}(t_{\rm init}) / m_{\rm e} c^2) \delta(t-t_{\rm init}) \delta(\gamma - \gamma_{\rm inj}) \\
        f_{\rm cont} (\dot{M}_{\rm sh}(t)/m_{\rm p})  \delta(\gamma - \gamma_{\rm inj}) \\
	\end{array} 
    \right.
    ,
\end{eqnarray}
where $\gamma_{\rm inj} \sim 1$, $m_{\rm p}$ is the proton mass and $\dot{M}_{\rm sh}(t)$ is the rate of the mass of the supernova (SN) ejecta swept-up by the PWN expansion (see section 2.2 of TK23).
$f_{\rm imp}$ and $f_{\rm cont}$ are the parameters which gives the particle injection rate to the stochastic acceleration process and should be smaller than unity.
TA17 found that the externally injected low energy particles can be accelerated stochastically to be the radio-emitting particles.
For the impulsive injection case $Q_{\rm ext}(\gamma, t) \propto f_{\rm imp}$, we consider that a fraction $f_{\rm imp}$ of the rotational energy at an early phase $t_{\rm init} (\ll$ 10 yr) is almost instantaneously converted into the low energy $e^{\pm}$ of $\gamma = \gamma_{\rm inj}$.
For the continuous injection case $Q_{\rm ext}(\gamma, t) \propto f_{\rm cont}$, we consider that a fraction $f_{\rm cont}$ of the swept-up ejecta neutrals is continuously injected to the stochastic acceleration process by photoionization inside the PWN and then a similar number of hadrons can also be injected and accelerated inside the nebula.
We ignore the acceleration and emission processes of the hadrons here, while we allow a fraction of the spin-down energy converted to ion acceleration, i.e., we set $\eta_{\rm T} + \eta_{\rm B} + \eta_{\rm e} < 1$ for the continuous injection case.
Note that the upper limit of the pressure (energy) of the accelerated hadrons is at most the sum of the other components inside the PWN, because they can affect the PWN expansion if the hadrons become dominant in pressure.
This upper limit is a requirement from the past studies of the dynamical and spectral evolution of PWNe \cite{Gelfand+09, Bucciantini+11}.

The other part of the model is the same as our previous studies, where synchrotron radiation and inverse Compton scattering off synchrotron radiation (SSC) and off the cosmic microwave background radiation (IC/CMB) from the expanding PWN inside the parent supernova remnant are calculated \cite{Tanaka&Takahara10, Tanaka&Takahara11, Tanaka&Takahara13a, Tanaka16, Tanaka&Asano17, Tanaka&Kashiyama23}.
The parameters of the model are those for the evolution of the supernova remnant $(E_{\rm SN}, M_{\rm SN}, n_{\rm ISM})$, those for the evolution of the central pulsar $(P_0, B_0, M_{\rm NS}, R_{\rm NS}, n)$, those for the injection of particles, magnetic fields, and turbulence from the pulsar $(p, \gamma_{\rm min}, \gamma_{\rm max}, \eta_{\rm e}, \eta_{\rm B}, \eta_{\rm T})$, and those for external particle injection ($f_{\rm imp}$ or $f_{\rm cont}$).

We apply the present model to the Crab Nebula in the next section \ref{sec:Results}.
The parameters of the parent supernova are set to $E_{\rm SN} = 10^{51}\ {\rm erg},\ M_{\rm ej} = 9.5 \ M_{\odot},$ and $n_{\rm ISM} = 0.1 \ {\rm cm^{-3}}$ \cite{MacAlpine&Satterfield08, Bucciantini+11} and those of the Crab pulsar are also set to $P_0 = 18.7$ msec, $B_0 = 3.35 \times 10^{12}$ G, $n = 2.51$, $M_{\rm NS} = 1.4~M_{\odot}$ and $R_{\rm NS} = 12$ km ($L_0 = 2.6 \times 10^{39}\ {\rm erg ~ s^{-1}},~\tau_0 = 1.1$ kyr).
The values of $\gamma_{\rm min}, \gamma_{\rm max}$ and $p$ are almost unique from previous studies (c.f. Refs. \cite{Kennel&Coroniti84b, Atoyan&Aharonian96}). 
The main parameter of the present study is only $\tau_{\rm acc}$ and we study the cases with $\tau_{\rm acc} = 1, 3, 10, 30, 100$ yr.
The other parameters, $(f_{\rm imp}, f_{\rm cont}), \eta_{\rm B}, \eta_{\rm T}, \eta_{\rm e}$, are allowed to change while they have already been constrained by our previous studies.
Both impulsive and continuous injection cases are studied.

\section{Results}\label{sec:Results}

%
\begin{figure}
    \begin{minipage}{0.5\textwidth}
    \centering
	\includegraphics[scale=0.55]{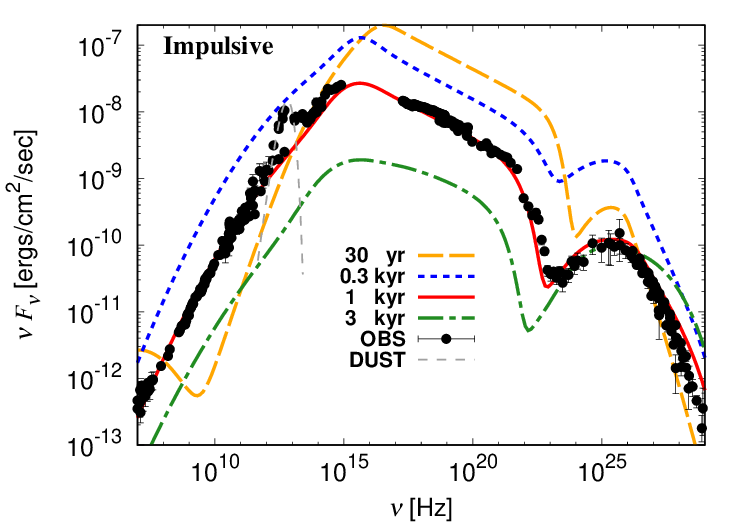} \\
    \subcaption{The spectral energy distribution}
	\includegraphics[scale=0.55]{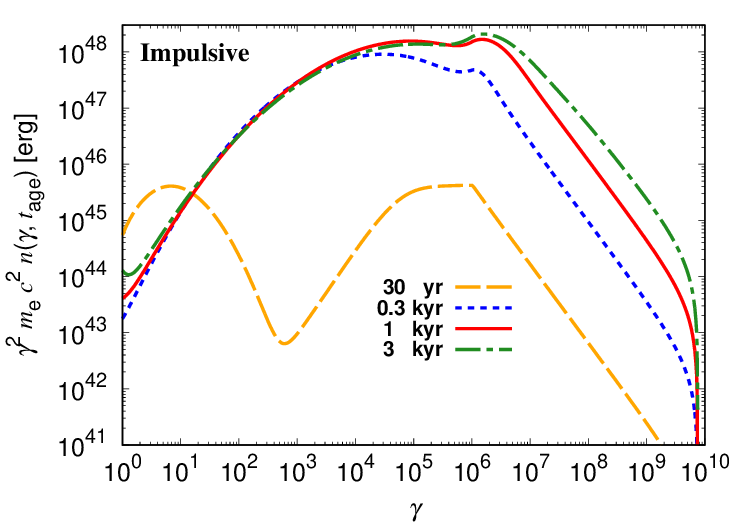}
    \subcaption{The particle energy spectra}
    \end{minipage}
    \begin{minipage}{0.5\textwidth}
    \centering
	\includegraphics[scale=0.55]{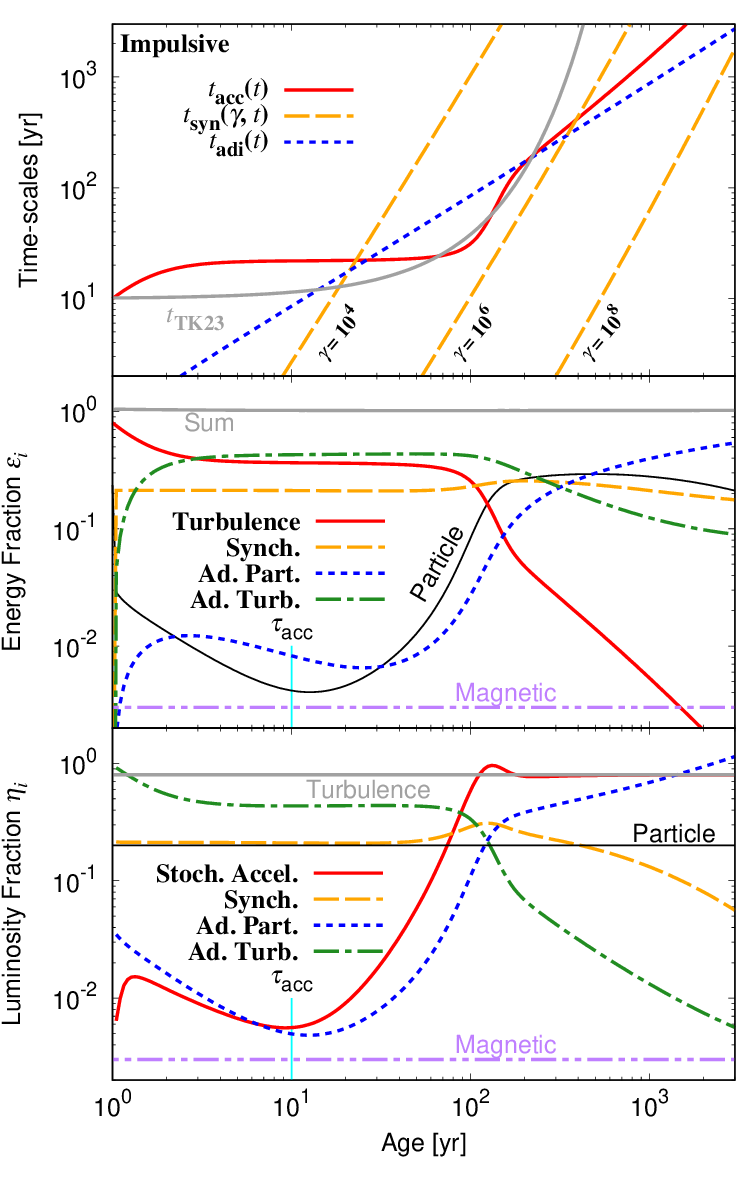}
    \subcaption{Evolution of the typical time-scales (top), the fractional energies (middle) and luminosities (bottom)}
    \end{minipage}
    \caption{
    The results for the impulsive injection case.
    (a) The spectral energy distribution and (b) the particle energy spectra (bottom) of the Crab Nebula are plotted for $t_{\rm age} =$ 30 yr (yellow dashed), 300 yr (blue dotted), 1 kyr (red solid) and 3 kyr (green dot-dashed), respectively.
    The black points on (a) are the observational data in radio \cite{Baldwin71, Baars+77, Macias-Perez+10, Weiland+11, PlanckCollab16, Ritacco+18}, in IR and optical \cite{Ney&Stein68, Grasdalen79, Green+04, Temim+06, Gomez+12, DeLooze19}, in X-rays \cite{Kuiper+01}, and in $\gamma$-rays \cite{Aharonian+06, Albert+08, Abdo+10, Amenomori+19, MAGICCollab20, Abeysekara+19, LHAASOCollabo21}.
    (c) Top panel: $t_{\rm acc}(t)$ (Eq. (\ref{eq:AccelerationTime}), red solid) and $t_{\rm adi}(t)$ (Eq. (\ref{eq:TurbulentEnergyEvolution}), blue dotted) evolve with time $t$, while the synchrotron cooling time-scales $t_{\rm syn}(\gamma,t)$ (yellow dashed) depends on both $t$ and $\gamma$, where $\gamma = 10^4, 10^6, 10^8$ are plotted.
    $t_{\rm TK23}(t)$ (definition in the text) is the acceleration time-scale adopted by TK23 (gray solid).
    (c) Middle panel: the fractional energy (per $E_{\rm rot}(t)$) of the particles $E_{\rm e}(t)$ (black solid), of the turbulence $E_{\rm T}(t)$ (red solid), of the magnetic field $E_{\rm B}(t) = \eta_{\rm B} E_{\rm rot}(t)$ (purple dot-dot dashed), of the particle coolings by synchrotron (yellow dashed) and adiabatic loss (blue dotted) $\int^t_0 d t'\int d \gamma |\dot{\gamma}_{\rm cool}(\gamma, t')| m_{\rm e} c^2 N(\gamma, t')$, of the adiabatic loss of the turbulent energy (green dot-dashed) $\int^t_0 E_{\rm e,T}(t')/t_{\rm adi}(t') d t'$ and of the sum of all of them (gray solid).
    (c) Bottom panel: the fractional luminosity (per $L_{\rm spin}(t)$) of stochastic acceleration backreaction (Eq. (\ref{eq:BackReaction}), red solid), synchrotron (yellow dashed) and adiabatic cooling rates for the particle (blue dotted) and the turbulence (green dot-dashed) energy.
    The energy injection rates $\eta_{\rm e,T,B}$ (black solid, gray solid, and purple dot-dot dashed) are also shown.
	The adopted parameters are summarized in Table \ref{tbl:Parameters}.
}
\label{fig:IMP}
\end{figure}


\begin{figure}
    \begin{minipage}{0.5\textwidth}
    \centering
	\includegraphics[scale=0.55]{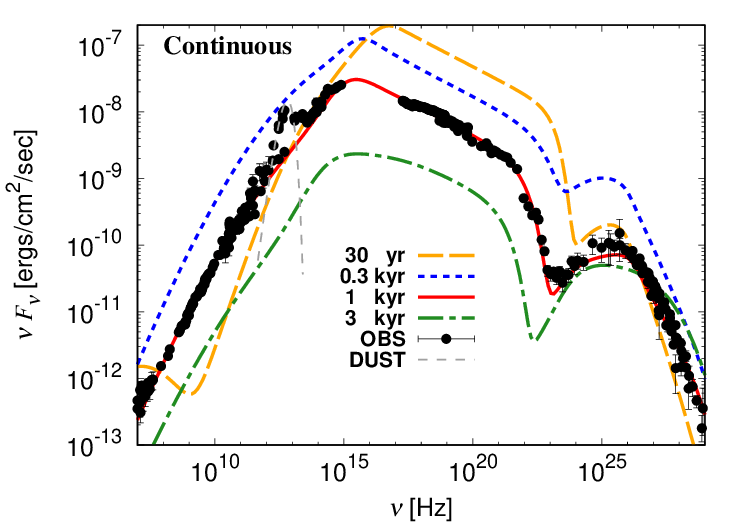} \\
    \subcaption{The spectral energy distribution}
	\includegraphics[scale=0.55]{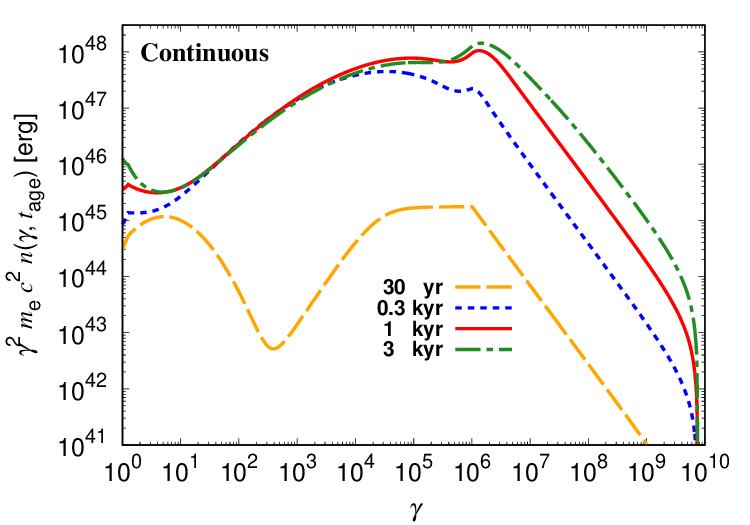}
    \subcaption{The particle energy spectra}
    \end{minipage}
    \begin{minipage}{0.5\textwidth}
    \centering
	\includegraphics[scale=0.55]{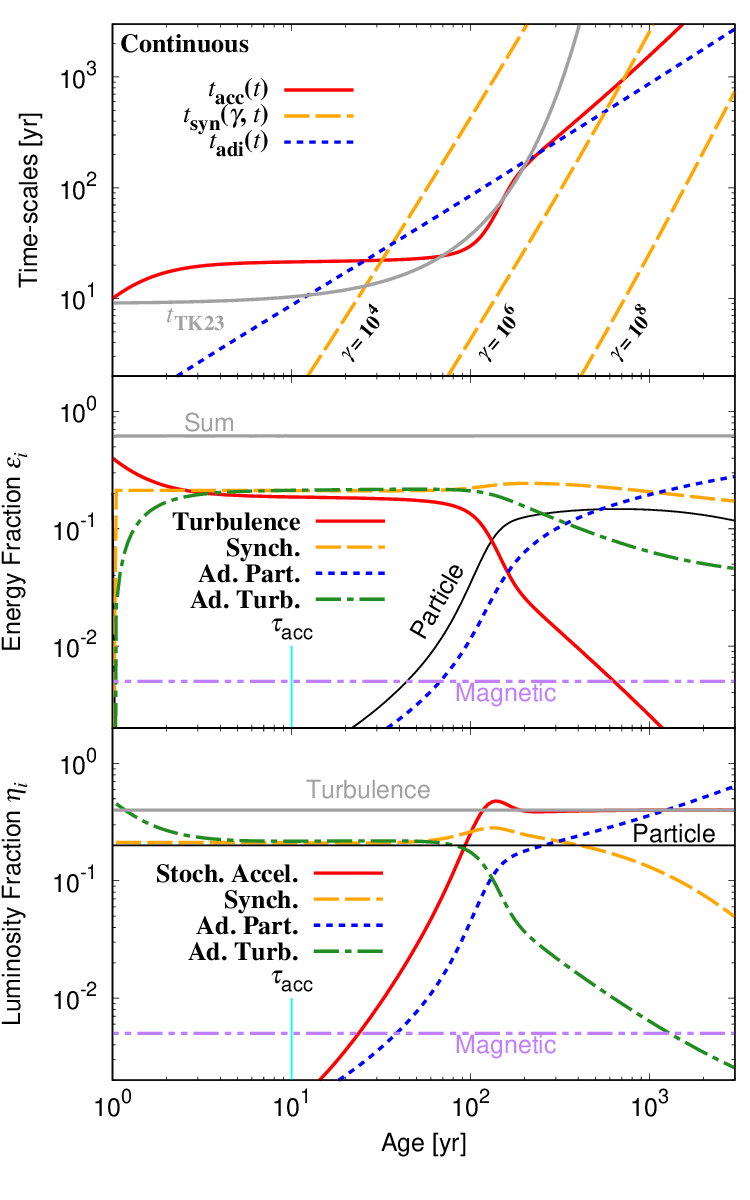}
    \subcaption{Evolution of the typical time-scales (top), the fractional energies (middle) and luminosities (bottom)}
    \end{minipage}
    \caption{
    The results for the continuous injection case.
    The adopted parameters are summarized in Table \ref{tbl:Parameters}.
}
\label{fig:CONT}
\end{figure}
Figs.\ref{fig:IMP} and \ref{fig:CONT} show evolution of spectra of (a) photons and (b) electrons for the impulsive and continuous injection cases, respectively.
We found that, for the both cases, the observed broadband spectrum of the Crab Nebula at an age of 1 kyr is reproduced with $\tau_{\rm acc} = 10$ yr and then it seems difficult to distinguish the impulsive injection from the continuous one with the one-zone spectral evolution model as the previous studies \cite{Tanaka&Asano17, Tanaka&Kashiyama23}.
Below, we just discuss the impulsive injection case in detail.
The used parameters are summarized in Table \ref{tbl:Parameters}.
Note that we obtained the parameters fitting by eye because the calculation is too heavy to make a statistical fit by doing a lot of runs.

The top panel of Fig. \ref{fig:IMP}(c) shows time evolution of the typical time-scales: the acceleration time $t_{\rm acc}(t)$ (red solid), the synchrotron cooling time $t_{\rm syn}(\gamma,t)$ (yellow dashed) and the adiabatic cooling time $t_{\rm adi}(t)$ (blue dotted).
The acceleration time-scale adopted by TK23 $t_{\rm TK23}$ (gray solid) is also plotted for reference, where $t_{\rm TK23}(t) = \hat{\tau}_{\rm acc} \exp(t/\hat{\tau}_{\rm turb})$ with $(\hat{\tau}_{\rm acc}, \hat{\tau}_{\rm turb}) = (10, 70)$ yr.
Only $t_{\rm syn}(\gamma, t) \propto \gamma^{-1} B_{\rm PWN}^{-2}(t)$ depends on $\gamma$ and rapidly increases with time because the magnetic field decreases with the PWN expansion.
The particles injected from the pulsar $Q_{\rm PSR}$ whose Lorentz factor is $\gamma \ge \gamma_{\rm min} = 10^6$ are not stochastically accelerated because $t_{\rm acc}(t)$ is always longer than the other time-scales for $\gamma \gtrsim \gamma_{\rm min}$.
The particles of $\gamma \lesssim 10^6$ are stochastically accelerated at a time when $t_{\rm acc}(t) < t_{\rm adi}(t)$ is satisfied, i.e., the stochastic acceleration starts at $t_{\rm age} \approx$ 25 yr and ends at $t_{\rm age} \approx$ 0.2 kyr.
The complicated behaviors of $t_{\rm acc}(t)$, such as an initial increase of a factor two around $t_{\rm age}  \sim$ yr, an exponential increase phase in $50 \lesssim t_{\rm age} \lesssim 130$ yr and a power-law increase phase after that, are not essential to reproduce the current observation, because the relatively simple $t_{\rm TK23}(t)$ still reproduce the observed flat radio spectrum \cite{Tanaka&Kashiyama23}.
Evolution of the turbulent energy $E_{\rm T}(t)$ reflects the complicated behaviors of $t_{\rm acc}(t)$ (Eq. (\ref{eq:TurbulentEnergyEvolution})).

The middle and bottom panels of Fig. \ref{fig:IMP}(c) are time evolution of each energy content per $E_{\rm rot}(t)$ and its time derivative (luminosity) per $L_{\rm spin}(t)$, respectively.
The adiabatic loss of the turbulent energy plays a role for the initial ($\sim$ yr) factor two increase of $t_{\rm acc}(t)$ (factor two decrease of $E_{\rm T}/E_{\rm rot}$, see dot-dashed green line in the bottom panel of Fig. \ref{fig:IMP}(c)).
$t_{\rm acc}(t) \propto E_{\rm rot}(t) / E_{\rm T}(t)$ is almost constant, i.e., $d E_{\rm T} / d t \sim \eta_{\rm T} L_{\rm spin}$, until the backreaction of the stochastic acceleration to the turbulence becomes significant $\le$ 0.1 kyr and after that $d E_{\rm T} / d t \sim (\delta E_{\rm T} / \delta t)_{\rm damp}$ (see Eq. (\ref{eq:BackReaction}) and red solid line in the bottom panel of Fig. \ref{fig:IMP}(c)).
Finally, $t_{\rm acc}(t)$ has a power-law dependence on time because $\eta_{\rm T} L_{\rm spin} \sim (\delta E_{\rm T} / \delta t)_{\rm damp}$.

We find that the current broadband observations can also be reproduced by taking $\tau_{\rm acc} =$ 1 and 3 yr with very similar sets of the parameters, while we do not find the solution with $\tau_{\rm acc} =$ 30 and 100 yr.
The results for $\tau_{\rm acc} =$ 3 and 30 yr are briefly summarized in Appendix \ref{app:DifferentTacc}.
We also studied the dependence on $t_{\rm init}$ by taking an order of magnitude small value of $t_{\rm init} =$ 0.1 yr and then the behavior of $t_{\rm age} \gtrsim$ a few years are almost the same as the results with $t_{\rm init} = 1$ yr shown in this section.

\section{Discussion \& Conclusions}\label{sec:Dis&Cons}

In this paper, the previous stochastic acceleration model is improved by considering the time evolution of the turbulent energy and then the total energy balance inside a PWN is maintained.
The turbulent energy supplied from the central pulsar is wasted by the backreaction from the stochastic particle acceleration and the adiabatic cooling according to the PWN expansion.
The model is applied to the Crab Nebula and reproduces the current broadband emission spectrum, especially the flat radio spectrum, although time evolution of the turbulent energy (diffusion coefficient) is different from our previous studies, where we assumed an exponential behavior of the diffusion coefficient (see the top panels of Figs. \ref{fig:IMP}(c) and \ref{fig:CONT}(c)).

The obtained acceleration time-scale of $\tau_{\rm acc} = 10$ yr or shorter (see Appendix \ref{app:DifferentTacc}) is consistent with TK23.
According to the discussion by TK23, the acceleration time-scale can be constrained by observing an early ($\lesssim$ a few decades) evolution of PWNe in radio band.
It is also difficult to distinguish the impulsive and continuous injection cases only from the present observational data \cite{Tanaka&Asano17, Tanaka&Kashiyama23}.
However, the continuous injection model would also accelerate ions which are simultaneously injected as electrons and then a characteristic neutrino signal would be expected.
We leave the study of the neutrino emission as a future study.

The impulsive and continuous injection models would have different spatial distributions of the radio-emitting particles.
For the impulsive injection case, the generated low energy $e^{\pm}$ are distributed throughout the dense PWN, and then the radio-emitting particles are expected to be almost uniformly distributed according to the PWN expansion.
For the continuous injection case, the particles are injected from the outer edge of the nebula, or more precisely, they are also injected from the supernova ejecta filaments, which penetrate the PWN due to the Rayleigh-Taylor instability \cite{Hester+96, Porth+14b}, so that we expect the radio-emitting particles to be distributed more in the outer part than near the center of the nebula.
In both cases, the spatial distribution of the radio-emitting particles should also be different from that of the X-ray-emitting particles injected from the central pulsar.
The spatial brightness profile also depends on the magnetic field structure and the spatial diffusion of the particles would also be important \cite{Porth+16, Ishizaki+17, Ishizaki+18, vanRensburg+20}.
Future observations and modeling of the spatial structure of PWNe will distinguish the origin of the radio-emitting particles.

The essence of the present study is the modeling of the acceleration time, i.e., Eq. \ref{eq:AccelerationTime}.
We consider the non-resonant acceleration with large-scale eddies of a compressible turbulence, which is characterized by the fact that the acceleration time `does not depend on the energy of the particles' ($D_{\gamma \gamma} \propto \gamma^2$) and `depends on the (spectrally integrated) energy density of the turbulence' \cite{Ptsuski88, Brunetti&Lazarian07}.
Since the present model is a one-zone model, we use the spatially integrated energy rather than the (local) energy density of the turbulence ($D_{\gamma \gamma} \propto E_{\rm T}(t)$).
The energy of the turbulence injected so far by the pulsar ($\eta_{\rm T} E_{\rm rot}(t)$) is introduced only to normalize the present turbulent energy inside the PWN.
In practice, the local physical quantities of the PWN plasma should be considered to determine the acceleration time, but it is not easy to define `waves' parameterized by mass density, temperature, etc. in the PWN.
The problem arises from the fact that the PWN consists of a relativistic electron-positron plasma with a `pure' non-thermal distribution, and then the accelerated particles scattered by the turbulence and the background plasma responsible for the turbulence are not easily distinguishable as usual.
The important conclusion obtained from the present study is that the requirement for $t_{\rm acc}(\gamma, t)$ in order to satisfy the radio observations is quite simple, i.e., the stochastic acceleration stops as the turbulent energy decreases (see the top panels of Figs. 1(c) and 2(c)).

The turbulence introduced to reproduce the broadband spectrum in this paper is not exactly the same as the turbulent magnetic field introduced to resolve the dynamical $\sigma$-problem \cite{Zrake&Arons17, Tanaka+18}.
The energy of the turbulence $E_{\rm T}(t)$ is the kinetic energy and is not the magnetic one because we do not include the synchrotron cooling by the turbulent magnetic field.
In addition, $E_{\rm T}(t)$ is directly supplied from the central pulsar but Tanaka et al. (2018) \cite{Tanaka+18} considered that the turbulent magnetic field is converted from the ordered magnetic field although the kinetic turbulence would convert the ordered magnetic field into the tangled one.
It is interesting that the turbulent energy $E_{\rm T}(t)$ is close to $E_{\rm B}(t)$ at an age of $\sim$ 1 kyr (see the middle panels of Figs. \ref{fig:IMP}(c) and \ref{fig:CONT}(c)), where the ratio of the turbulent to ordered magnetic field is expected to be order unity from the polarization observation \cite{Mizuno+23}. 
The turbulence inside PWNe would be a plausible solution to the long-standing problems of the PWN physics and then we should relate and combine them in the future studies.

\section*{Acknowledgment}

S. J. T. would like to thank S. Kisaka, K. Murase, Y. Ohira, and K. Nishiwaki for useful discussion.
The authors would like to thank the anonymous referee for helpful comments.
This work is supported by JPJS Bilateral Program, Grant No. JPJSBP120229940 (SJT) and by JSPS KAKENHI 24H01816 (SJT), 21J01450 (WI).
S. J. T. would like to thank the Sumitomo Foundation, and the Research Foundation For Opto-Science and Technology for support.

\appendix

\section{Results for different acceleration time-scales}\label{app:DifferentTacc}

%
\begin{figure}
\begin{center}
	\includegraphics[scale=0.55]{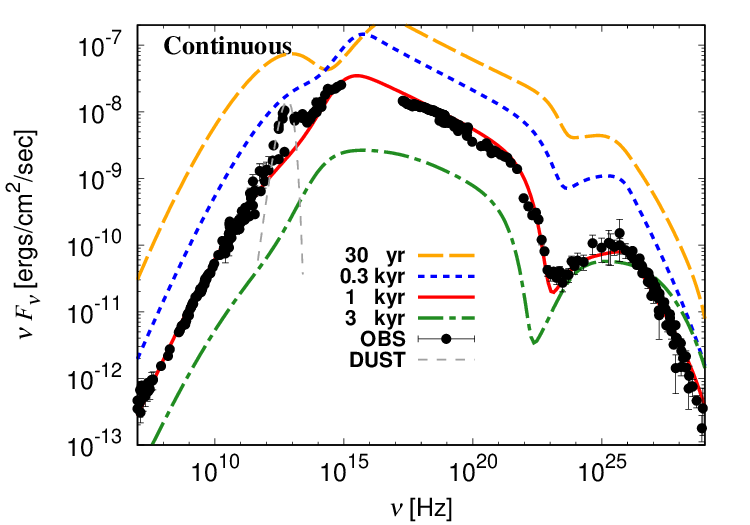}
	\includegraphics[scale=0.55]{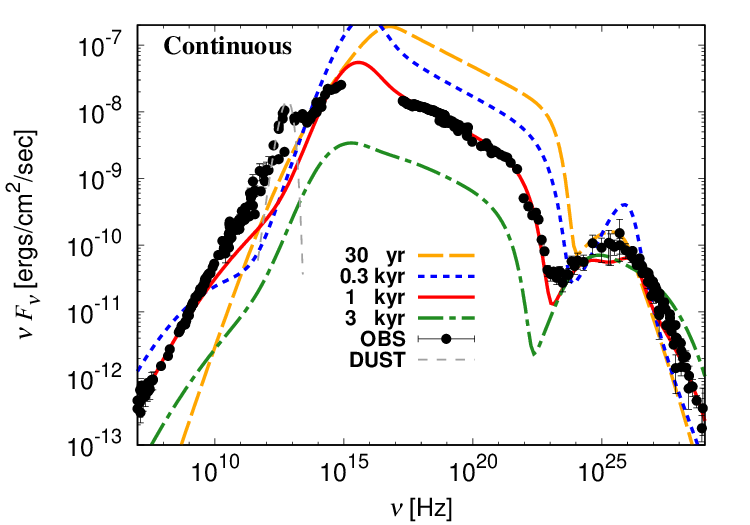}
\end{center}
\caption{
    The spectral energy distribution for $\tau_{\rm acc} = 3$ yr (left) and 30 yr (right), respectively.
}\label{fig:CONT_SED_3_30}
\begin{center}
	\includegraphics[scale=0.55]{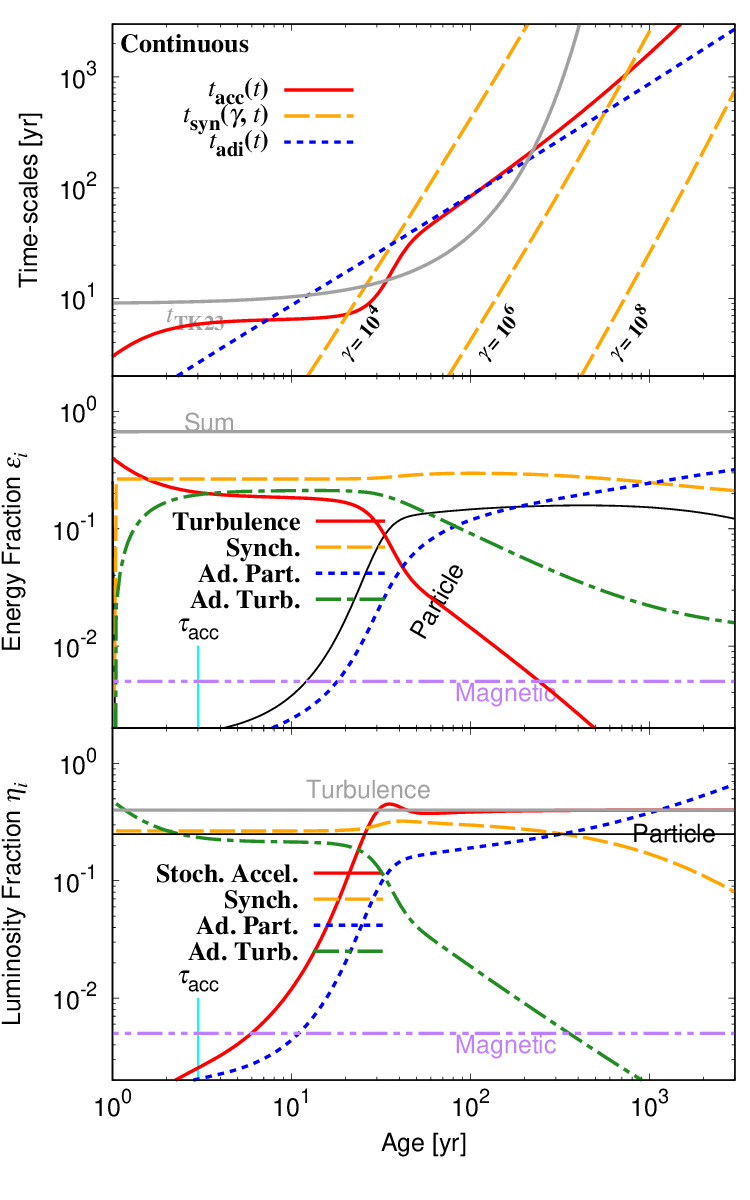}
	\includegraphics[scale=0.55]{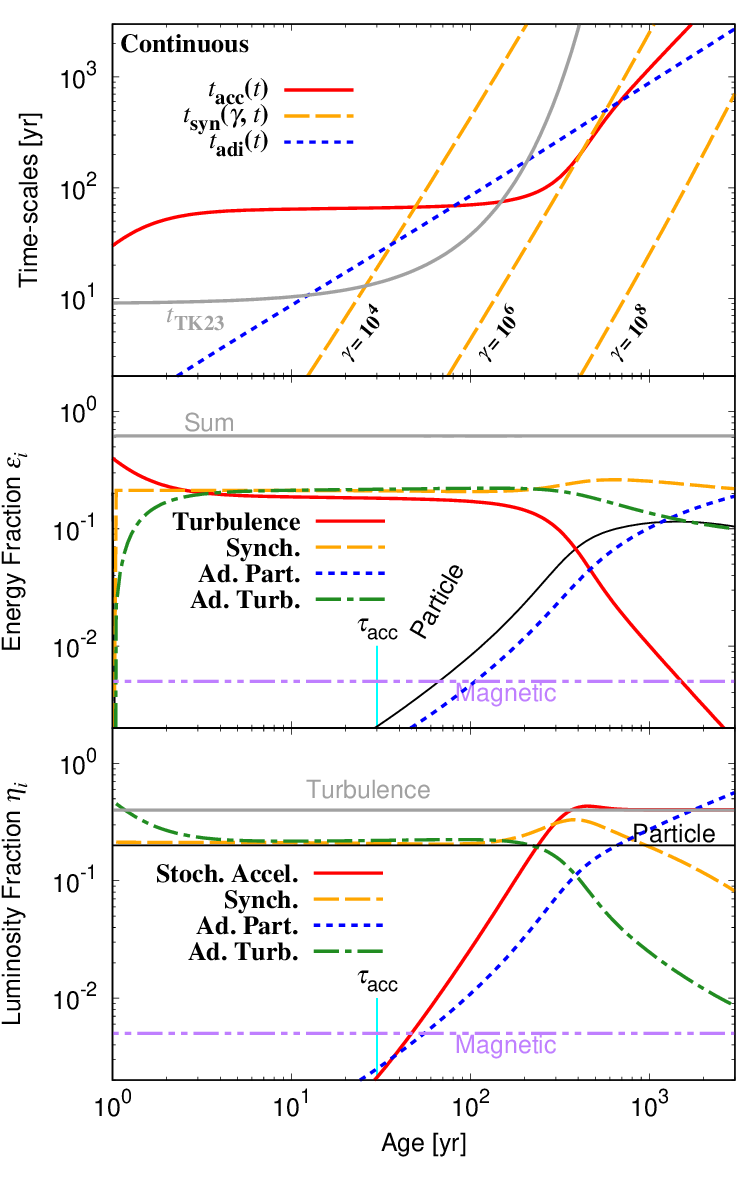}
\end{center}
\caption{
    Evolution of the typical time-scales (top), the fractional energies (middle) and luminosities (bottom) for $\tau_{\rm acc} = 3$ yr (left) and 30 yr (right).
}\label{fig:CONT_EVOL_3_30}
\end{figure}

Figs. \ref{fig:CONT_SED_3_30} and \ref{fig:CONT_EVOL_3_30} show evolution of the spectral energy distribution and the typical time-scales, the fractional energies (middle) and luminosities (bottom) for $\tau_{\rm acc} = 3$ yr (left) and 30 yr (right), respectively.
The results for the continuous injection case are shown.
The parameters are almost the same as the case of $\tau_{\rm acc} =$ 10 yr (Fig. \ref{fig:CONT}).
We do not obtain the flat radio spectrum for $\tau_{\rm acc} =$ 30 yr while the radio observations can be reproduced for $\tau_{\rm acc} =$ 3 yr. 
The turbulent energy is not fully converted into the energy of the radio-emitting particles for $\tau_{\rm acc} =$ 30 yr.
We conclude that $\tau_{\rm acc} \le 10$ yr is required to reproduce the current radio observations of the Crab Nebula.


\bibliographystyle{ptephy}
\bibliography{draft}

\end{document}